\documentclass[prl,twocolumn,superscriptaddress,showpacs,floatfix]{revtex4}
\usepackage{graphicx,amsmath,amssymb,bm}

\newcommand{\fmi}{\, \text{fm}^{-1}}
\newcommand{\mev}{\, \text{MeV}}
\newcommand{\vlowk}{V_{{\rm low}\,k}}

\begin{document}

\title{Three-body forces and the limit of oxygen isotopes}

\author{Takaharu Otsuka}
\affiliation{Department of Physics,
University of Tokyo, Hongo, Tokyo 113-0033, Japan}
\affiliation{Center for Nuclear Study,
University of Tokyo, Hongo, Tokyo 113-0033, Japan}
\affiliation{National Superconducting Cyclotron Laboratory,
Michigan State University, East Lansing, MI, 48824, USA}

\author{Toshio Suzuki}
\affiliation{Department of Physics, College of Humanities and Sciences,
Nihon University, Sakurajosui 3, Tokyo 156-8550, Japan}

\author{Jason D.\ Holt}
\affiliation{TRIUMF, 4004 Wesbrook Mall, Vancouver, BC, V6T 2A3, Canada}

\author{Achim Schwenk}
\affiliation{TRIUMF, 4004 Wesbrook Mall, Vancouver, BC, V6T 2A3, Canada}

\author{Yoshinori Akaishi}
\affiliation{RIKEN Nishina Center, Hirosawa, Wako-shi, Saitama
351-0198, Japan}

%\date{\today}

\begin{abstract}
The limit of neutron-rich nuclei, the neutron drip-line, evolves
regularly from light to medium-mass nuclei except for a striking
anomaly in the oxygen isotopes. This anomaly is not reproduced in
shell-model calculations derived from microscopic two-nucleon
forces. Here, we present the first microscopic explanation of the
oxygen anomaly based on three-nucleon forces that have been
established in few-body systems. This leads to repulsive
contributions to the interactions among excess neutrons that change
the location of the neutron drip-line from $^{28}$O to the
experimentally observed $^{24}$O. Since the mechanism is robust and
general, our findings impact the prediction of the most neutron-rich
nuclei and the synthesis of heavy elements in neutron-rich
environments.
\end{abstract}


\maketitle

\begin{figure}[t]
\begin{center}
\includegraphics[width=0.45\textwidth,clip=]{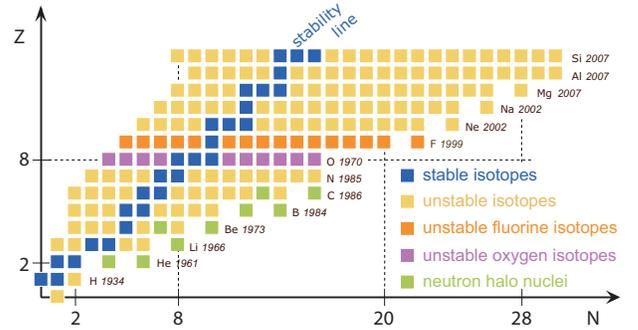}
\end{center}
\caption{Stable and unstable nuclei with 
$Z \leqslant 14$ and neutron number $N$~\cite{masstable}.
The oxygen anomaly in the location of the neutron drip-line
is highlighted. Element names and years of discovery of the
most neutron-rich nuclei are given. The axis numbers indicate the
conventional magic numbers.
\label{Fig1}}
\end{figure}

One of the central challenges of nuclear physics is to develop a
unified description of all nuclei created in the laboratory and the
cosmos based on the underlying forces between neutrons and protons
(nucleons). This involves understanding the sequences of isotopes in
the nuclear chart, Fig.~\ref{Fig1}, from the limits of proton-rich
nuclei to the neutron drip-line. These limits have been established
experimentally up to oxygen with proton number $Z$=8. Mapping out the
neutron drip-line for larger $Z$~\cite{Baumann} and exploring
unexpected structures in neutron-rich nuclei are a current frontier in
the physics of rare isotopes. The years of discovery in
Fig.~\ref{Fig1} highlight the tremendous advances made over the last
decade.

Figure~\ref{Fig1} shows that the neutron drip-line evolves regularly
with increasing proton number, with an odd-even bound-unbound pattern
due to neutron halos and pairing effects. The only known anomalous
behavior is present in the oxygen isotopes, where the drip-line is
strikingly close to the stability line~\cite{Oxygen}. Already in the
fluorine isotopes, with one more proton, the drip-line is back to the
regular trend~\cite{Sakurai}. In this Letter, we discuss this puzzle
and show that three-body forces are necessary to explain why 
$^{24}$O~\cite{Janssens,N16} is the heaviest
oxygen isotope.

Three-nucleon (3N) forces were introduced in the pioneering work of
Fujita and Miyazawa (FM)~\cite{FM} and arise because nucleons are
composite particles.
The FM 3N mechanism is due to one nucleon virtually exciting
a second nucleon to the $\Delta(1232 \mev)$ resonance, which is
de-excited by scattering off a third nucleon, see Fig.~\ref{diagrams}(e).

Three-nucleon interactions arise naturally in chiral effective field
theory (EFT)~\cite{chiral}, which provides a systematic basis for
nuclear forces, where nucleons interact via pion exchanges and
shorter-range contact interactions.
The resulting nuclear forces are
organized in a systematic expansion from leading to successively
higher orders, and include the $\Delta$ excitation as the dominant
part of the leading 3N forces~\cite{chiral}. The quantitative role of
3N interactions has been highlighted in recent ab-initio calculations
of light nuclei with $A=N+Z \leqslant 12$~\cite{GFMC,NCSM}.

We first discuss why the oxygen anomaly is not reproduced in
shell-model calculations derived from microscopic NN forces. This can
be understood starting from the stable $^{16}$O and adding neutrons
into single-particle orbitals (with standard quantum numbers $nlj$)
above the $^{16}$O core. We will show that correlations do not change
this intuitive picture.
Starting from $^{16}$O, neutrons
first fill the $0d_{5/2}$ orbitals, with a closed subshell
configuration at $^{22}$O ($N=14$), then the $1s_{1/2}$ orbitals at
$^{24}$O ($N=16$), and finally the $0d_{3/2}$ orbitals at $^{28}$O
($N=20$). For simplicity, we will drop the $n$ label in the following.

\begin{figure}[t]
\begin{center}
\includegraphics[width=0.46\textwidth,clip=]{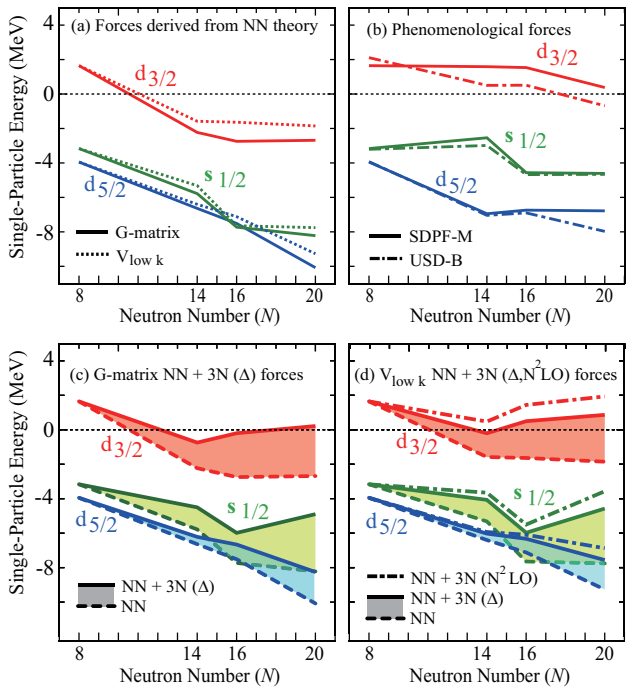}
\end{center}
\caption{Single-particle energies of the neutron $d_{5/2}$, $s_{1/2}$
and $d_{3/2}$ orbitals measured from the energy of $^{16}$O as a
function of neutron number $N$. (a) SPE
calculated from a $G$ matrix and from
low-momentum interactions $\vlowk$. (b) SPE obtained
from the phenomenological forces SDPF-M~\cite{sdpfm} and
USD-B~\cite{usdb}. (c,d) SPE including contributions from 3N forces
due to $\Delta$
excitations and chiral EFT 3N interactions at N$^2$LO~\cite{3Nfit}.
The changes due to 3N forces
based on $\Delta$ excitations are highlighted by the shaded
areas.\label{Fig2}}
\end{figure}

\begin{figure}[t]
\begin{center}
\includegraphics[width=0.32\textwidth,clip=]{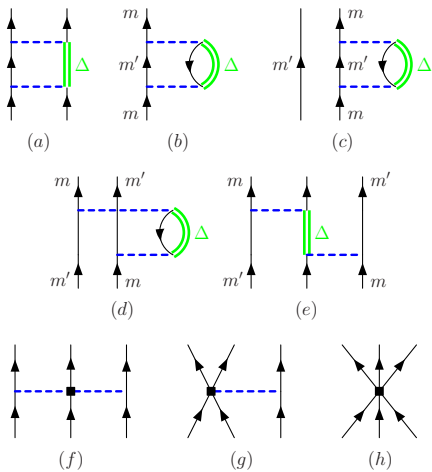}
\end{center}
\caption{Processes involving 3N contributions. The external lines
are valence neutrons. The dashed and thick lines denote pions and
$\Delta$ excitations, respectively. Nucleon-hole lines are indicated
by downward arrows. The leading chiral 3N forces include the long-range
two-pion-exchange parts, diagram~(f), which take into account the
excitation to a $\Delta$ and other resonances, plus shorter-range
one-pion exchange, diagram~(g), and 3N contact interactions,
diagram~(h).\label{diagrams}}
\end{figure}

In Fig.~\ref{Fig2}, we show the single-particle energies (SPE) of the
neutron $d_{5/2}$, $s_{1/2}$ and $d_{3/2}$ orbitals at subshell
closures $N=8$, 14, 16, and 20. The evolution of the SPE is due
to interactions as neutrons are added. For the SPE based on NN forces
in Fig.~\ref{Fig2}~(a), the $d_{3/2}$ orbital decreases rapidly as
neutrons occupy the $d_{5/2}$ orbital, and remains well-bound from
$N=14$ on. This leads to bound oxygen isotopes out to $N=20$ and puts
the neutron drip-line incorrectly at $^{28}$O. This result appears to
depend only weakly on the renormalization method or the NN interaction
used. We demonstrate this by showing SPE calculated in the $G$ matrix
formalism~\cite{Gmatrix}, which sums particle-particle ladders, and
based on low-momentum interactions $\vlowk$~\cite{Vlowk} obtained from
chiral NN interactions at next-to-next-to-next-to-leading order
(N$^3$LO)~\cite{N3LO} using the renormalization group. Both
calculations include core polarization effects perturbatively
(including diagram
Fig.~\ref{diagrams}~(d) with the $\Delta$ replaced by a nucleon and
all other second-order diagrams) and start from empirical SPE~\cite{sdpfm}
in $^{17}$O. The empirical SPEs contain effects from
the core and its excitations,  
including effects due to 3N forces.

We next show in
Fig.~\ref{Fig2}~(b) the SPE obtained from the phenomenological forces
SDPF-M~\cite{sdpfm} and USD-B~\cite{usdb} that have been fit to
reproduce experimental binding energies and spectra. This shows a
striking difference compared to Fig.~\ref{Fig2}~(a): As neutrons
occupy the $d_{5/2}$ orbital, with $N$ evolving from 8 to 14, the
$d_{3/2}$ orbital remains almost at the same energy and is not
well-bound out to $N=20$. The dominant differences between 
Figs.~\ref{Fig2}~(a) and (b) 
can be traced to the two-body monopole
components, which determine the average interaction between two
orbitals. The monopole components of a general
two-body interaction $V$ are given by an angular average over all possible
orientations of the two nucleons in orbitals $lj$ and
$l'j'$~\cite{Bansal},
\begin{equation}
V^{\rm mono}_{j,j'} = \sum_{m,m'} \langle j m \, j' m' | V | j m \, 
j' m' \rangle \, \bigl/ \sum_{m,m'} 1 \,,
\end{equation}
where the sum over magnetic quantum numbers $m$ and $m'$ can be 
restricted by antisymmetry (see \cite{RMP,mono} for details).
The SPE of the orbital $j$ is
effectively shifted by $V^{\rm mono}_{j,j'}$ multiplied by the
occupation number of the orbital $j'$. This leads to the change in
the SPE and determines shell structure and the location of
the drip-line~\cite{RMP,mono,magic,tensor}.

The comparison of Figs.~\ref{Fig2}~(a) and~(b) suggests that the
monopole interaction between the $d_{3/2}$ and $d_{5/2}$ orbitals
obtained from NN theories is too attractive, and that the oxygen anomaly
can be solved by additional 
repulsive contributions to the two-neutron monopole
components, which approximately cancel the average NN attraction on
the $d_{3/2}$ orbital. With extensive studies based on NN forces, it
is unlikely that such a distinct property would have been missed, and
it has been argued that 3N forces may be important for the
monopole components~\cite{zuker}.

Next, we show that 3N forces among two valence neutrons and one
nucleon in the $^{16}$O core give rise to repulsive monopole
interactions between the valence neutrons. While the contributions of
the FM 3N force to other quantities can be different, the shell-model
configurations composed of valence neutrons probe the long-range parts
of 3N forces. The repulsive nature of this 3N mechanism can be
understood based on the Pauli exclusion principle.
Figure~\ref{diagrams}~(a) depicts the leading contribution
to NN forces due to the excitation of a $\Delta$, induced by the
exchange of pions with another nucleon. Because this is a second-order
perturbation, its contribution to the energy and to the two-neutron
monopole components has to be attractive. This is part of the
attractive $d_{3/2}$-$d_{5/2}$ monopole component obtained from NN
forces.

In nuclei, the process of Fig.~\ref{diagrams}~(a) leads to a change of
the SPE of the $j, m$ orbital due to the excitation of a core nucleon
to a $\Delta$, as illustrated in Fig.~\ref{diagrams}~(b) 
where the initial valence neutron is virtually
excited to another $j', m'$ orbital. As discussed, this lowers the
energy of the $j, m$ orbital and thus increases its binding. However,
in nuclei this process is forbidden by the Pauli exclusion principle,
if another neutron occupies the same orbital $j', m'$, as shown in
Fig.~\ref{diagrams}~(c). The corresponding contribution must then be
subtracted from the SPE change due to Fig.~\ref{diagrams}~(b). This is
taken into account by the inclusion of the exchange diagram,
Fig.~\ref{diagrams}~(d), where the neutrons in the intermediate state
have been exchanged and this leads to the exchange of the final (or
initial) orbital labels $j, m$ and $j', m'$. Because this process
reflects a cancellation of the lowering of the SPE, the contribution
from Fig.~\ref{diagrams}~(d) has to be repulsive for two neutrons.
Finally, we can rewrite Fig.~\ref{diagrams}~(d) as the FM 3N force of
Fig.~\ref{diagrams}~(e), where the middle nucleon is summed over core
nucleons. The importance of the cancellation between
Figs.~\ref{diagrams}~(a) and~(e) was recognized for nuclear matter
in Ref.~\cite{Brown}.

\begin{figure*}[t]
\vspace*{-18mm}
\begin{center}
\includegraphics[width=0.86\textwidth,clip=]{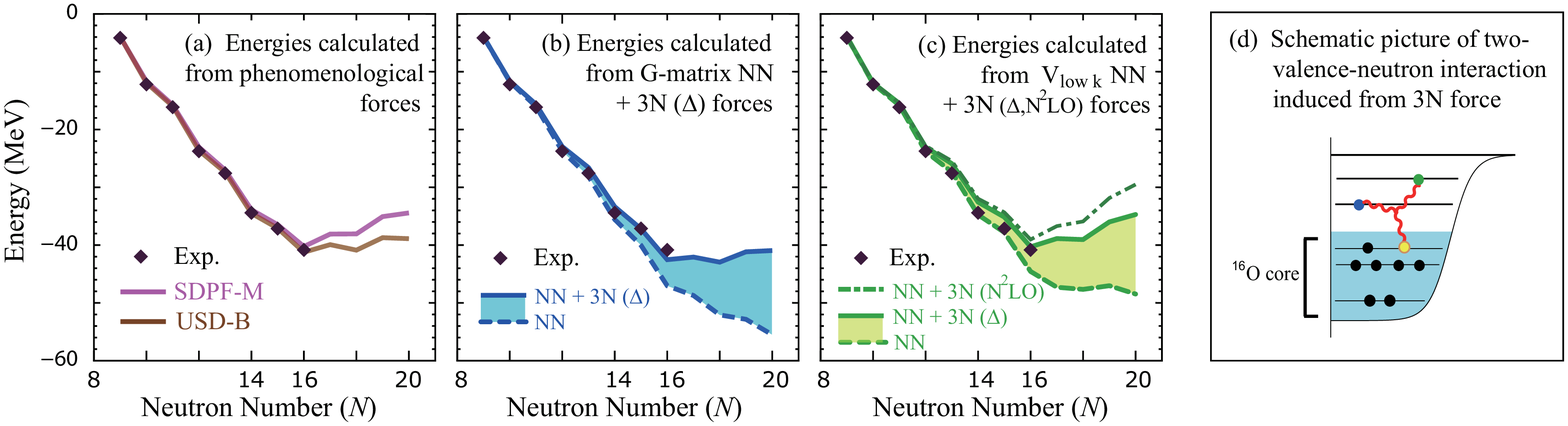}
\end{center}
\caption{Ground-state energies of oxygen isotopes measured from
$^{16}$O, including experimental values of the bound $^{16-24}$O.
Energies obtained from
(a)~phenomenological forces SDPF-M~\cite{sdpfm} and
USD-B~\cite{usdb}, (b)~a $G$ matrix and including FM 3N forces due
to $\Delta$ excitations, and (c)~from low-momentum interactions
$\vlowk$ and including chiral EFT 3N interactions at N$^2$LO as well
as only due to $\Delta$ excitations~\cite{3Nfit}. The changes due to
3N forces based on $\Delta$ excitations are highlighted by the
shaded areas. (d)~Schematic illustration of a two-valence-neutron
interaction generated by 3N forces with a nucleon in the $^{16}$O
core.\label{gs}}
\end{figure*}

The process in Fig.~\ref{diagrams}~(d) corresponds to a 
two-valence-neutron monopole interaction, schematically illustrated
in Fig.~\ref{gs}~(d). The resulting SPE evolution
is shown in Fig.~\ref{Fig2}~(c) for the $G$ matrix
formalism, where a standard pion-N-$\Delta$
coupling~\cite{green} was used and all 3N diagrams 
of the same order as Fig.~\ref{diagrams}~(d)
%obtained from antisymmetrization 
are included. We observe that the repulsive FM 3N
contributions become significant with increasing $N$ and the resulting
SPE structure is similar to that of phenomenological forces, where the
$d_{3/2}$ orbital remains high. 
Next, we calculate the SPE
from chiral low-momentum interactions $\vlowk$, including the changes
due to the leading (N$^2$LO) 3N forces in chiral EFT~\cite{chiral3N},
see Figs.~\ref{diagrams}~(f)--(h).
%%% (at the monopole level).
We consider also the SPE where 3N-force contributions are only
due to $\Delta$ excitations~\cite{Deltafull}.
%Next, we consider the SPE calculated
%from chiral low-momentum interactions $\vlowk$ and include the changes 
%due to the leading (N$^2$LO) 3N forces in chiral EFT~\cite{chiral3N},
%see Figs.~\ref{diagrams}~(f)--(h), as well as due to $\Delta$
%excitations~\cite{Deltafull} (at the monopole level).
The leading chiral 3N forces include the
long-range two-pion-exchange part, Fig.~\ref{diagrams}~(f), which
takes into account the excitation to a $\Delta$ and other resonances,
plus shorter-range 3N interactions, Figs.~\ref{diagrams}~(g) and~(h),
that have been constrained in few-nucleon systems~\cite{3Nfit}. 
The resulting SPE in Fig.~\ref{Fig2}~(d) demonstrate that the long-range
contributions due to $\Delta$ excitations dominate the changes in the
SPE evolution and the effects of shorter-range 3N interactions are
smaller. We point out that 3N forces play a key role for
the magic number $N=14$ between $d_{5/2}$ and $s_{1/2}$~\cite{N14}, 
and that they enlarge the $N=16$ gap between $s_{1/2}$ and $d_{3/2}$ 
~\cite{N16}.

The contributions from Figs.~\ref{diagrams}~(f)--(h) (plus all exchange
terms) to the monopole components take into account the normal-ordered
two-body parts of 3N forces, where one of the nucleons is summed over
all nucleons in the core. This is also motivated by recent
coupled-cluster calculations~\cite{CC3N}, where residual 3N forces
between three valence states were found to be small. 
In addition, the effects of 3N forces
among three valence neutrons should be generally weaker due to the
Pauli principle.

Finally, we take into account many-body correlations by
diagonalization in the valence space.
The resulting ground-state energies of the
oxygen isotopes are presented in Fig.~\ref{gs}. Figure~\ref{gs}~(a)
(based on phenomenological forces) implies that
many-body correlations do not change our picture developed from the
SPE: The energy decreases to $N=16$, but the $d_{3/2}$ neutrons added
out to $N=20$ remain unbound. Figures~\ref{gs}~(b) and~(c) give the
energies derived from NN forces, using a $G$ matrix or low-momentum
interactions $\vlowk$, and including two-valence-neutron interactions
due to 3N forces at the monopole level~\cite{multipole}.  
For all results based
on NN forces, the energy decreases to $N=20$ and the neutron drip-line
is incorrectly located at $^{28}$O. The changes due to 3N forces based
on $\Delta$ excitations are highlighted in Fig.~\ref{gs}~(b)
and~(c). This leads to a better agreement with the experimental
energies and to a kink at $N=16$, which is further strengthened by
shorter-range 3N forces, and for Fig.~\ref{gs}~(c) leads to the
neutron drip-line at $^{24}$O.

The same 3N forces lead to repulsion in neutron matter~\cite{neutmatt}.
Our results are also consistent with
early shell-model explorations with 3N forces up to $^{21}$O, where a
small repulsive effect as in Figs.~\ref{gs}~(b) and~(c) was
found~\cite{PollsRath}. Because the formation of a halo is unrealistic
for the $d_{3/2}$ orbital and $s_{1/2}$ is well bound (see
Fig.~\ref{Fig2}~(b)), it seems unlikely that the ground states beyond
$N=16$ become bound by including the coupling to the continuum.  
This is consistent with Ref.~\cite{tsukiGSM}. 
We plan to study 3N-force effects on unbound states in the future
using the methods of Refs.~\cite{tsukiGSM,CCSM}.
Fluorine isotopes have one more proton than oxygen, and NN forces,
primarily the tensor part, with this proton provide more binding to
the valence neutrons~\cite{tensor,F_utsuno}. This valence
proton-neutron effect is absent in the oxygen isotopes, making the
repulsive 3N mechanism visible. Important directions for future work
are to include the presented 3N contributions in coupled-cluster
calculations~\cite{Hagen} and in density-functional calculations, 
to systematically explore the effect over the full range of
the nuclear chart.

In summary, we have presented a robust 3N mechanism that provides
repulsive monopole interactions between valence neutrons. Using
microscopic NN and 3N forces as well as known SPE, our shell-model
calculations naturally explain why $^{24}$O is the heaviest oxygen
isotope. The changes due to 3N forces are amplified and testable in
neutron-rich nuclei and are expected to play a crucial role for matter
at the extremes.

We thank S.\ Bogner, R.\ Furnstahl, A.\ Nogga and T.\ Nakamura for
useful discussions. This work was supported in part by grants-in-aid
for Scientific Research~(A) 20244022 and~(C) 18540290, by the JSPS
Core-to-Core program EFES, by EMMI and NSERC.
TRIUMF receives funding via a contribution through the
National Research Council Canada. Part of the numerical calculations
have been performed at the JSC, J\"ulich, Germany.

\end{document}